# A lattice calculation of the branching ratio for $\overline{B} \to K^*\gamma$

UKQCD collaboration, presented by Hugh Shanahan [a]

[a]Department of Physics and Astronomy, University of Edinburgh, Edinburgh EH9 3JZ, Scotland

Preliminary results are presented on the calculation of the relevant form factor for the radiative decay $\overline{B} \to K^*\gamma$. We find that the form factor is weakly dependent on the spectator quark mass. We compare the results obtained for a full chiral extrapolation and where this independence is assumed. Both results are found to be statistically consistent with the experimental data.

## 1. Introduction

The decay $\overline{B} \to K^*\gamma$ is unique in being simultaneously of interest to experimentalists, and the of supersymmetry and lattice field theory communities. As the quark decay $b \to s\gamma$ cannot occur at tree-level, it is sensitive to a range of parameters in the Standard Model, including $|V_{ts}V_{tb}^*|$ and $m_t$. It is also sensitive to new physics from two Higgs' doublets models [1], supersymmetry [2], and even supergravity [3]. While there is currently debate on how much could be determined from this decay for these models, it is clear that $b \to s\gamma$ is a good probe of possible new physics, and that accurate theoretical predictions are essential.

$\overline{B} \to K^*\gamma$ is an excellent candidate to explore this quark decay experimentally, as it has a clean signature. The CLEO collaboration have published a branching ratio for $\overline{B} \to K^*\gamma$ of $(4.5 \pm 1.5 \pm 0.9) \times 10^{-5}$ [4], albeit with quite low statistics. However, unlike $b \to s\gamma$, there are large soft-gluonic contributions to this decay, and different models have given a wide range of different answers. It is clear then that non-perturbative methods, and lattice field theory in particular offer a way ahead.

Following the notation of Grinstein, Springer and Wise [5], the branching ratio can be evaluated in the leading log approximation from the matrix element

$$\frac{eG_F m_b}{2\sqrt{2}\pi^2} C_7(m_b) V_{tb} V_{ts}^* \eta^{\mu*} \langle K^*|\overline{s}\sigma_{\mu\nu} q^\nu b_R|B\rangle$$

where $C_7(m_b)$ is calculated perturbatively and $q^\mu, \eta^\mu$ are respectively the momentum and polarisation of the photon. Bernard, Hsieh and Soni [6] demonstrated the feasibility of a lattice calculation of $\langle B|\overline{b}\sigma_{\mu\nu} q^\nu s|K^*\rangle$, which can expressed using three form factors $T_i(q^2)$ multiplying three possible Lorentz structures $C_\mu^i$:

$$\langle B|\overline{b}\sigma_{\mu\nu}q^\nu s|K^*\rangle = \sum_{i=1}^{3} C_\mu^i T_i(q^2). \quad (1)$$

For on-shell photons, $q^2=0$ and the contribution from the third form factor vanishes, furthermore

$$T_2(q^2{=}0) = -iT_1(q^2{=}0). \quad (2)$$

Hence, the branching ratio can be expressed as a product of a number of parameters which are quite accurately known, $C_7(m_b)^2$, which depends on $m_t$, and $T_1(q^2{=}0)^2$. This paper concentrates on evaluating $T_1(q^2{=}0)$.

## 2. Computational details

The analysis for this calculation is based on 60 quenched configurations for a $24^3 \times 48$ lattice at $\beta = 6.2$, generated on the Meiko i860 Computing Surface based at Edinburgh. From string tension measurements [7], the lattice spacing $a$ was determined to be $2.73(5)$ GeV. The algorithms used in the gauge field generation and calculating matrix elements of the form $\langle A|\overline{q}_1\Gamma q_2|B\rangle$, where $|A\rangle$ and $|B\rangle$ are meson states, are explained in [7] and [8].

The quark propagators are evaluated using the $\mathcal{O}(a)$ improved ("clover") fermionic action, with periodic boundary conditions. To simulate the spectator (light) quarks, the propagators are evaluated at $\kappa_l = 0.14144, 0.14226$ and $0.14262$, whose pole masses are approximately that of the strange quark, given that $\kappa_s^{phys} = 0.1419(1)$ [9]. For the other quark in the $K^*$ state, the first two $\kappa$ values are used as they straddle the physical strange quark mass. Heavy quark propagators are computed at $\kappa_h = 0.129$ and $0.121$, with pole masses of approximately 1.9 and 2.5 GeV. The operator $\bar{h}\sigma_{\mu\nu}s$ is improved to $\mathcal{O}(a)$.

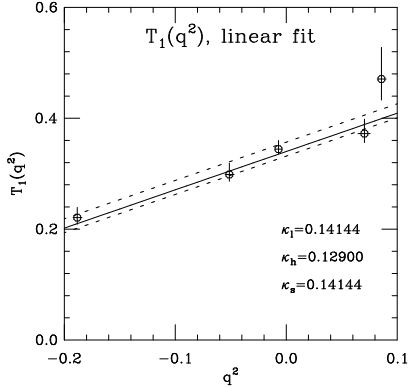

Figure 1. $T_1(q^2)$, with a linear fit. The dotted lines represent the 68% confidence levels of the fit at $q^2=0$.

Gauge invariant smearing is used for the extended pseudoscalar field to improve the signal. We apply time–reversal symmetry in the evaluation of $\sum_\epsilon \epsilon_\rho \langle V(k,\epsilon)|\bar{s}\sigma_{\mu\nu}b|P(p)\rangle$ for the same reason. The matching of the lattice operator with the continuum [10] has a negligible contribution and is not considered here.

$T_1$ is evaluated from the above by considering different components of the relation

$$4(k^\alpha p^\beta - p^\alpha k^\beta)T_1(q^2) = \varepsilon^{\alpha\beta\rho\mu}\sum_\epsilon \epsilon_\rho \langle V(k,\epsilon)|\bar{s}\sigma_{\mu\nu}b|P(p)\rangle q^\nu. \quad (3)$$

The resulting summed data is averaged on both sides of the lattice. A correlated fit to a constant is carried out for timeslices 11 to 13. The statistical analysis for this and all the subsequent fits of this paper is based on the bootstrap algorithm, using 250 subsamples.

The form factor is evaluated for seventeen different values for the spatial momentum $\vec{q}$, ranging in magnitude from 0 to $2\pi/(12a)$, and for $\vec{p} = (\pi/(12a), 0, 0)$ and $(0, 0, 0)$. For this choice of momenta and $\kappa$ values, the range of $q^2$ always includes zero and is small in comparison to the scale expected from the vector meson dominance. Hence, for each combination of $\kappa_l, \kappa_s$ and $\kappa_h$, $T_1(k_l, k_s, h_h; q^2)$ is linearly interpolated to $q^2=0$. An example of this fit is shown in figure 1.

The variation of $T_1$ with respect to the quark masses is now considered.

## 3. Extrapolation to physical mass scales

The data set $T_1(\kappa_l, \kappa_s, \kappa_h; q^2=0)$ is used to derive the form factor at the appropriate mass scales. Given the high statistics of this calculation, $T_1(\kappa_l, \kappa_s, \kappa_h; q^2=0)$ can be extrapolated to the chiral limit. It is fitted to the function

$$T_1(\kappa_l, \kappa_s, \kappa_h; q^2=0) = T_1(\kappa_{crit}, \kappa_s, \kappa_h; q^2=0) + \Delta(\kappa_s, \kappa_h)m_{q\,light}, \quad (4)$$

where $m_{q\,light}$ is the pole mass of the light quark. An example of such a fit is in figure 2. The result of this fit, $T_1(\kappa_{crit}, \kappa_s, \kappa_h; q^2=0)$, is linearly interpolated to $\kappa_s^{phys}$. The two remaining values of the form factor are extrapolated to the $B$ mass scale with the function

$$T_1(\kappa_{crit}, \kappa_h; q^2=0) = T_1^{static} + \frac{A}{M_P}, \quad (5)$$

giving a result of $T_1(q^2=0) = 0.15^{+12}_{-14}$. We note that the chiral extrapolation of the form factor exhibits a very weak dependence on the light quark pole mass. If we assume the data is indeed independent of the light quark mass and fit to a constant, we find the $\chi^2$ per degree of freedom to be of the same order as fitting the data to a linear function. If the data from this fit is continued to the physical regimes one finds $T_1(q^2=0) = 0.15^{+5}_{-4}$, significantly reducing the statistical errors and maintaining the central values. Both extrapolations to the $B$ mass scale are shown in figure 3.

For comparison, the branching ratio of $\overline{B} \to K^*\gamma$ from CLEO is converted into an experimental value for the form factor, $T_1^{expt}$. Using the parameters from the particle data book, and taking $m_b = 4.39\,\text{GeV}$, $T_1^{expt}$ is found to be 0.23(6), 0.21(5) and 0.19(5) for $m_t = 100$, 150 and 200 GeV. These results are statistically consistent with those found from our analysis.

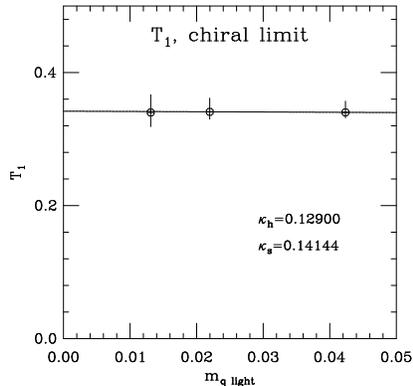

Figure 2. Chiral extrapolation of $T_1(q^2=0)$. $m_{q\ light}$ is the lattice pole mass.

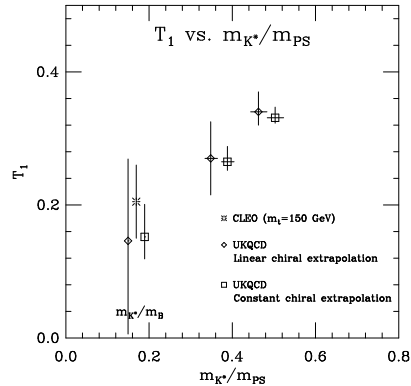

Figure 3. Extrapolation of $T_1(q^2=0)$ to $m_B$. The fitted points are displaced for clarity.

## 4. Conclusions

The calculation of matrix elements from three point functions on the lattice is rapidly approaching maturity. In our case, the derivation of the matrix element $\langle V|\bar{s}\sigma_{\mu\nu}h|P\rangle$ in the quenched approximation presents no conceptual difficulties and gives a good signal. The question remains whether extrapolating this data to the chiral limit and the appropriate heavy mass scale is valid, as the former is susceptible to large systematic errors due to finite volume effects and the latter does not have a clear theoretical justification for the ansatz used.

In the first case, it is very encouraging that the form factor $T_1$ has no, or at least a very weak dependence on the light quark mass, indicating a stability in the data. To address the second issue, we are currently repeating the calculation on the same lattice, using two new heavy quark masses at $\kappa_h = 0.125$ and $0.133$. Furthermore, we plan to repeat the calculation on a $16^3 \times 48$ lattice at $\beta = 6.0$, generating the heavy quark propagator with a hopping parameter expansion. This will allow us to explore a large number of heavy quark masses. It shall also check for any systematic errors due to the lattice spacing.

With our current data, we shall repeat the analysis presented by Soni at this conference [11], where the form factor $T_2$ is extracted. This can be used as a further determination of the systematic errors.

## 5. Acknowledgments

We would like to thank A. Soni, C. Bernard, A. El-Khadra and C. Allton for useful discussions on this topic. The Meiko $i860$ Computing Surface is supported by SERC grant $GR/G32779$, Meiko Limited, and the University of Edinburgh. This work was supported by SERC grant $GR/H01069$. Personal support of HS by the Wingate Foundation is gratefully acknowledged.

## REFERENCES

1. T.G. Rizzo, Phys. Rev. D **38**, 820 (1988).
2. M.A. Diaz, Phys. Lett. B**304**, 278 (1993).
3. J.L. Lopez, D.V. Nanopoulos and G.T. Park, Phys. Rev. D**48**, 974 (1993).
4. CLEO Collaboration, R. Ammar et al., Phys. Rev. Lett. **71**, 674 (1993).
5. B. Grinstein, R. Springer and M.B. Wise, Nucl.Phys B**339**, 269 (1990).
6. C.W. Bernard, P.F. Hsieh and A. Soni, Nucl. Phys. B**26**, 347 (1992).
7. UKQCD collaboration, C. Allton et al., Nucl. Phys. B**407**, 331 (1993).
8. UKQCD collaboration, S.P. Booth et al., Edinburgh preprint 93/525, hep–lat 9308019, submitted for publication in Phys. Rev. Lett.
9. UKQCD collaboration, C. Allton et al., Edinburgh preprint 93/524, hep–lat 9309002, to appear in Phys. Rev. D**49**, 1 (1994).
10. A. Borelli, C. Pittori, R. Frezzotti, and E. Gabrielli, CERN–TH–6587–92, (1992).
11. A. Soni, these proceedings.